\documentclass[sigconf]{acmart}

\AtBeginDocument{%
  }

\usepackage{graphicx}
\usepackage{multirow}
\usepackage{tabularx}

\copyrightyear{2026}
\acmYear{2026}
\makeatletter
\def\@ACM@copyright@check@cc{}
\makeatother
\setcopyright{cc}
\setcctype{by}
\acmConference[WiPSCE 2026]{The 20th WiPSCE Conference on Primary and Secondary Computing Education Research}{March 11--13, 2026}{Aachen, Germany}
\acmBooktitle{The 20th WiPSCE Conference on Primary and Secondary Computing Education Research (WiPSCE 2026), March 11--13, 2026, Aachen, Germany}
\acmDOI{10.1145/3801749.3801763}
\acmISBN{979-8-4007-2502-9/2026/03}

\begin{document}

\title{Make it SewSimple: Navigating UK Curriculum and Classroom Practice in Secondary Computing Education with E-textiles}
\title[E-Textiles in Secondary Computing Education]{Make it SewSimple: Navigating UK Curriculum and Classroom Practice in Secondary Computing Education with E-textiles}

\author{Yifan Feng}
\authornote{Both authors contributed equally to this research.}
\orcid{0009-0009-9589-7589}
\affiliation{%
\department{UCL Knowledge Lab \& UCL Interaction Centre}
\institution{University College London} 
\city{London} 
\country{UK}}
\additionalaffiliation{%
\department{Creative Computing Institute, }
  \institution{University of the Arts London}
  \city{London}
  \country{United Kingdom}}
\email{yifanfeng4@acm.org} 

\author{Hanlin Zhang}
\authornotemark[1]
\orcid{0009-0008-8156-7290}
\affiliation{%
\department{UCL Knowledge Lab}
\institution{University College London} 
\city{London} 
\country{UK}}
\email{hanlin.zhang.20@ucl.ac.uk}

\author{Yishan Du}
\orcid{0009-0000-1116-659X}
\affiliation{%
\department{UCL Knowledge Lab}
\institution{University College London} 
\city{London} 
\country{UK}}
\email{yishan.du.24@ucl.ac.uk}

\author{Weihong Tang}
\orcid{0009-0007-2096-9061}
\affiliation{%
\department{UCL Knowledge Lab}
  \institution{University College London}
  \city{London}
  \country{UK}}
  \email{weihong.tang.24@alumni.ucl.ac.uk}

\author{Jennifer A. Rode}
\orcid{0000-0002-3511-3975}
\affiliation{%
\department{UCL Knowledge Lab}
\institution{University College London} 
\city{London} 
\country{UK}}
\email{jen@acm.org}

\author{Bea Wohl}
\orcid{0000-0002-5969-9210}
\affiliation{%
\department{Creative Computing Institute}
  \institution{University of the Arts London}
  \city{London}
  \country{UK}}
\email{b.wohl@arts.ac.uk}

\renewcommand{\shortauthors}{Feng et al.}

\begin{abstract}
This paper explores the potential of integrating e-textiles as part of the approach to delivering computing in UK secondary schools. As one of the few UK-based exploratory studies of teachers’ experiences, it investigates how e-textile platforms such as the ``SewSimple'' maker kit and the BBC micro:bit can be incorporated into Key Stage 3 computing education (ages 11-14), taking into account both English national curriculum requirements and the realities of classroom practice. Our research question is: How do teachers perceive the potential of including e-textiles as part of computing education in English secondary schools? In summary, our research contributes to secondary computing education in three ways. First, we examine teachers’ direct, cross-disciplinarity experiences in two participatory design workshops using a newly designed e-textile platform and extend the limited discussion on supporting the BBC micro:bit in e-textile education. Second, we specifically identify opportunities, barriers, and challenges across three dimensions: school planning, national curriculum guidance, and practical e-textile implementation. Third, we offer insights into best practices for supporting maker technology adoption and pedagogical practices within existing institutional structures to maximize students' benefits for secondary school computing education. 
\end{abstract}


\maketitle

\section{Introduction}
Ubiquitous computing systems have gained prevalence in CS education due to rising societal demands to teach young generations that ``information is everywhere'', enabling them to make informed decisions \citep{przybylla2017social}. As a result, physical computing, a concept originally from art and design, was introduced to challenge stereotypes in computing education \citep{igoe2004physical, grillenberger2023and}, emphasizing ``motivation, creativity, and constructionist learning'' to teach core computing knowledge, despite being overlooked in earlier K12 classroom practice \citep{przybylla2017nature, Przybylla2018thesis}.  In the past decades, a growing number of studies have highlighted the benefits of integrating e-textiles, a specific type of physical computing platform, in pre-college computing education to gain students' interest and improve the engagement in STEM \citep{kafai2019stitching, kafai2014crafts, guimerans2024textiles} and foster collective creativity \citep{chandrashekar2025product}, particularly for gender minorities \citep{weibert2014sewing, rode2024exploring, buechley2008lilypad, fey2022anywear} and other underrepresented student groups \citep{fields2023supporting, kafai2014electronic, buechley2013textile}, including one case study focusing on the UK context \citep{coulter2023textiles}. While prior work \citep{Przybylla2018thesis} presents practical pedagogical guidelines for introducing physical computing in classroom teaching, little attention has been given to the unique material affordances of e-textiles. Moreover, much of the existing e-textiles research has focused primarily on student learners’ experiences, with several research by Fields and colleagues (e.g., \citep{fields2023supporting, fields2017teaching}) that has explored training computing teachers with e-textile knowledge through professional development programs called ``Exploring Computer Science (ECS)'' in the US and highlighted key elements—such as supporting teachers’ social–emotional development—that are critical for sustainable e-textile education in computing. Building on this foundation, our study offers one of the few UK-based exploratory studies of teachers' experiences and to investigate how e-textile platforms such as the ``SewSimple'' maker kit and the BBC micro:bit can be integrated into Key Stage 3 computing education (students aged 11-14), taking into account both national curriculum requirements and the everyday realities of classroom practice. This paper explores the research gap in how teachers envision the use of e-textiles within the national curriculum framework, which could lead to greater uptake and engagement in community skills and subjects among underrepresented groups in computing education.  

We report insights from two participatory design workshops conducted with eight UK-based teachers to explore perceived barriers, opportunities and challenges for bringing e-textile platforms into English computing classrooms. Through six hands-on co-design activities focused on technology and instructional design evaluation, our goal was to better understand how the unique environment of UK secondary computing education constrains the equity goals of e-textiles within computing culture. Our research question is: \textit{How do teachers perceive the potential of including e-textiles as part of computing education in English secondary schools?} 

In summary, this research contributes to computing education in three ways. First, we examine teachers’ direct, cross-disciplinary experiences with scenario-based learning using a newly designed e-textile platform by our team and extend the limited discussion on supporting the BBC micro:bit—a physical computing technology backed by the UK government—in e-textile education. Second, we identify opportunities, barriers, and challenges. Third, we offer insights into best practices for supporting e-textile adoption and pedagogical practices within existing institutional structures to maximize students' benefits for CS education.

\section{Related Work}
\subsection{BBC micro:bit and CS Education in the UK}
The BBC launched the micro:bit initiative in 2016 in the United Kingdom, and early evaluations highlighted both the motivational appeal of the device and the novelty of engaging students in physical computing in computing education \citep{rogers2017bbc, austin2020bbc, hodges2020physical, raspberrypi_report} and computer literacy \citep{Schmidt2016ComputerLiteracy}, reporting that students associated the it being able to ``create cool stuff'' \citep{sentance2017a}. Some studies revealed that the device itself does not guarantee effective learning; its value is mediated by pedagogical choices and teachers’ pedagogical approaches, which rely on adequate scaffolding and professional development \citep{sentance2017b, videnovik2018}. Other studies explored the potential of the device for interdisciplinary and deeper disciplinary learning \citep{videnovik2018, tzagkaraki2023} and designed several research toolkits to support pedagogical practices \citep{underwood2024tangible, palin2024microdata, underwood2025toolkit, underwood2024microcode, feng2025sewsimple}. While BBC micro:bit intended to ``inspire every child to create their best digital future'', especially with a goal of ``widening participation around gender and disadvantaged groups'' \citep[p.67]{austin2020bbc}, scholars  highlight the importance of implementing sensitive teaching methods to complement the use of such learning technologies. For example, \citet{videnovik2018} showed that with targeted encouragement, girls were not only able to match boys in programming tasks but in some cases excelled, underscoring that disparities reflect opportunity structures rather than innate differences. In line with these findings, studies further confirm that girls typically report lower self-efficacy in programming tasks and less confidence in their abilities compared to boys, highlighting the need for targeted interventions to build equitable confidence and motivation in computing education \citep{kallia2018boys, aivaloglou2019early}. Equally, \citet{tyren2018} provided practical recommendations;  introducing computational concepts through analog activities, student-friendly terminology, and preparing for technical pitfalls. A survey of 405 U.K. students and teachers using micro:bit usersshows that 86\% of students find computing more enjoyable, and girls are 70\% more likely to consider computing as a subject after exposure to the device \citep{austin2020bbc}. \citet{tzagkaraki2023} demonstrated that a micro:bit-based robotics sequence significantly enhanced students’ self-efficacy and scientific career motivation.

Important limitations remain, particularly regarding inclusivity and transparency. Employing the ``Computational Making'' framework \citep{rode2015computational}, in \citep{rode2024exploring, rode2024spools}, they show how hardware constraints (the lack of sewable connectors) restricted creativity and aesthetics, undermining inclusivity in textile-based projects. Similarly, \citet{cederqvist2022} found that students often treated the device as a ``black box'', limiting conceptual gains. This reinforces the need for teachers to frame micro:bit activities not only as hands-on making but also as opportunities for students to engage with concepts such as feedback systems, materials properties, and the dual nature of technological artifacts. Building on these insights, our study explores the potential for the classroom implementation of an e-textile platform and BBC micro:bit.

\subsection{E-textile and CS Education}
Several studies have examined teachers' experiences introducing e-textiles to pre-college computing education. Teachers generally found that making and crafting practices made teaching core computing concepts more engaging and transformative \citep{nakajima2019teachers, nakajima2020lighting}. E-textile emphasis on personalization and problem-solving helped teachers understand students’ individual ideas, and explore students’ expertise, collaboration, and peer learning \citep{fields2017teaching}.  E-textiles can help make computing relevant to all learners while fostering more welcoming and inclusive classroom environments \citep{shaw2020leveraging}. By focusing on strengthening school–community \citep{fields2023supporting} and school–home connections \citep{nakajima2019teachers} through creating tangible and computational artifacts, teachers can design learning activities that allow students to link their computing practice to tactile and embodied form of knowledge such as cultural identity and lived experiences \citep{scott2015culturally}, and build an authentic and positive teacher-student relationships for equitable computing education \citep{nakajima2019teachers, shaw2020leveraging}. 

Despite these benefits, research also reveals emerging challenges in teaching e-textiles within computing classes. For example, \citet{shaw2020leveraging} found that teachers' primary concern was a lack of pedagogical approaches and instructional materials that balance computing concepts with students' individual needs and interests. While several studies have proposed examples of e-textile teaching content (e.g., \citep{buechley2007towards, ioannou2025learning, litts2017stitching}), this work has largely focused on informal learning settings such as workshops, and less is known about how these approaches translate to formal classroom instruction. 

Regarding teachers' skill development for teaching e-textiles, \citet{fields2023supporting} highlighted the challenge of integrating aesthetically driven e-textiles and physical computing into teachers' existing knowledge and practice. They also identified several supports essential for effective professional development, including stronger organizational backing, access to material and technical resources, and the cultivation of cross-disciplinary teacher communities and socio-emotional support to sustain ongoing teacher training \citep{fields2023supporting}. These suggestions focus on United States CS teachers, leaving questions about the challenges and supports needed to introduce e-textiles into UK-based context. Building on these insights, our work addresses this gap by exploring context-sensitive recommendations for e-textiles in teaching computing subjects that reflect the culture and pedagogical practices.

\subsection{Issues with Teaching Computing in the UK} 
The current computing curriculum \citep{govuk_k3_computing} in England, introduced in 2014, was a shift from the previous ICT framework, aiming for a deeper understanding of the principles rather than focusing solely on using software \citep{wohl2025learning, de2014game, royalsociety_report} to meet the needs of the digital economy \citep{wohl2017future}. However, implementing the new computing curriculum posed challenges for teaching. The switch from ICT to computing required teachers to move from delivering basic skills to teaching programming, algorithms, and computational thinking \citep{wohl2025learning}. Many primary teachers had little or no background in computing, while secondary ICT teachers often lacked the depth of expertise required, making teacher retraining essential \citep{boulton2018implementing}. Pedagogical change has also been important, as teaching programming demands a more student-centered, problem-solving approach. Teachers faced several barriers, including limited time for professional development, workload pressures, and school-level support \citep{de2014game}.  Communities of practice, regional hubs, and master teachers have successfully built networks to share resources and improve confidence; the reach of these initiatives remains uneven, and the sustainability of such support is uncertain \citep{boulton2018implementing}. Meanwhile, motivation of students is also a concern, with teachers reporting student disengagement, particularly among girls, and frustration when students give up quickly \citep{sentance2017computing}. These challenges highlight that while the curriculum is ambitious in vision, its success depends on systemic investment in teacher capacity, resources, and evidence-informed practices to ensure equitable and effective delivery.

Some researchers have suggested that computational thinking and computing should be introduced earlier and in more playful, engaging ways—such as through games, robots, and hands-on activities—to build stronger conceptual foundations \citep{wohl2025learning}. One approach is the use of tools like the BBC micro:bit, which allows students to experiment with physical computing and see the immediate, tangible results of their code \citep{kvavsvsayova2022experience}. This approach can make abstract concepts such as variables, loops, and algorithms more concrete, supporting students’ problem-solving skills and creativity \citep{austin2020bbc}. One educational framework designed explicitly with e-textiles in mind, computational making, created a framework to support maker education and computing education to address this wider set of skills \citep{rode2015computational}, and we feel this needs to be taken up in future work. Such strategies can also foster a sense of ownership over learning, which may encourage wider participation and help address gender disparities in computing. As such, this research will explore challenges of teaching computing using ``SewSimple'' and micro:bit.

\section{Method}
\begin{figure}[t]
  \centering
  \includegraphics[width=0.35\columnwidth]{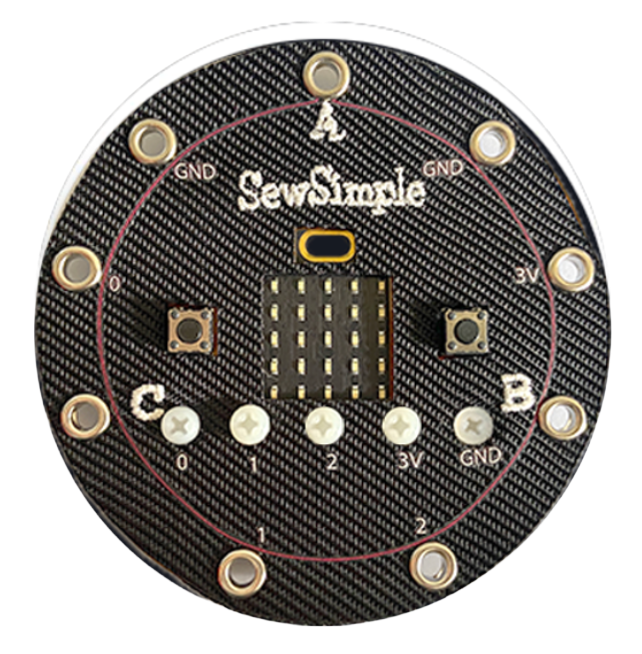}
  \caption{The ``SewSimple (V1.1)'' Tool}
  \Description{}
  \label{fig:sewsimple_maker}
\end{figure}

This study investigated UK-based teachers’ perspectives on e-textiles using a newly designed maker kit by our team (See Figure. \ref{fig:sewsimple_maker}), the ``SewSimple'' \citep{feng2025sewsimple} and BBC micro:bit for secondary school computing education. Building on \citet{musaeus2024bringing}'s discussion, we conducted two participatory design workshops with eight teachers (current or newly qualified). While all participants were encouraged to take part in both workshops, only two (P1 and P4) were able to (see Table \ref{tab:workshop_participants}). Both workshops were held in person in London at University College London in Summer 2025, with ethics approval obtained in advance (\#REC2071 \& \#RKEESC020\_25\_01). Participant was compensated with a £20 Amazon voucher on the completion of each workshop as an acknowledgment of their time.

\subsection{Participants}
We recruited teachers who were familiar with the Key Stage 3 (KS3) national curriculum (students aged 11-14) and had at least one year of teaching experience at public or private schools at either the upper primary/lower secondary level in England (detailed information see Table \ref{tab:workshop_participants}). Participants were initially recruited via the university networks (e.g., directors of PGCE/teacher training programs, social media, and personal contacts). As e-textiles require multidisciplinary skillsets and cross-subject collaboration \citep{kafai2014crafts}, we sought to include not only computing teachers but also teachers from related subjects (e.g., Design \& Technology, Art \& Design) as discussed in prior work \citep{coulter2023textiles} to enrich our data collection. In total, five computing teachers and three art \& design teachers participated in the two design workshops. 

\subsection{Workshop Structure}
The overarching goal was to gather teachers’ in-depth feedback on e-textiles in computing education, in relation to the current national curriculum guidance defined by the UK government. 

\subsubsection{Workshop 1}
The first workshop
focused on identifying the relevance of e-textiles, potential challenges, and the resources required to integrate them into current classroom practice through three activities. To support a contextualized understanding and discussion, inspired by the cultural probe technique \citep{perry2018designing, gaver1999design}, we introduced teachers to a novel e-textile teaching/maker kit called ``SewSimple''. Participants engaged in a series of exploratory tasks, including touching and feeling electronic and non-electronic materials (e.g., felt, colored pencils), inspecting example e-textile artifacts, and practicing a basic activity of lighting an external LED using ``SewSimple'' and a BBC micro:bit to simulate an introductory teaching experience. This was followed immediately by a structured reflection task. For the reflection, we provided a two-column whiteboard (``Usability challenges'' / ``Support needed''). Participants wrote their ideas on sticky notes and placed them in the relevant column. In doing so, they clarified the usability challenge and the corresponding forms of support they would find most useful.

Afterwards, the five participants were divided into two groups to examine a sample e-textile project and collaboratively map its alignment with the National Curriculum for England (KS3) Computing and Design \& Technology goals. Groups were composed to balance gender, subject expertise (Computing v.s. Art \& Design/Textiles), and experience; however, exact parity was not possible with five participants. In this activity, participants worked on a pre-prepared whiteboard: a left-hand reference column listed the KS3 computing goals \citep{govuk_k3_computing}, with three mapping columns to the right—``Aligned elements'', ``Challenges to alignment'', and ``Suggestions for alignment''. In small groups, participants first located the relevant KS3 goal in the reference column. They then used the ``Aligned elements'' column to note which feature(s) of the sample e-textile project evidenced that alignment, and recorded strengths, misalignments, and proposed improvements in the relevant columns. Building on this, each teacher completed an individual reflection on the provided board, writing their own ideas on sticky notes in three sections: ``Challenge'', ``Supports'', and ``Resources'' to articulate what they considered necessary for effective curriculum alignment. The final portion brought the two groups back together to reflect on their group findings from the workshop and teaching experiences.

\subsubsection{Workshop 2}
The second workshop 
extended these discussions, aiming to co-envision an e-textile teaching framework based on the UK national curriculum's guidance through three design activities. Prior work highlights the benefits of scenario-based learning (SBL) in fostering computational thinking through inquiry-based activities and educational technologies \citep{zitouniatis2023teaching, feng2025sewsimple, norton2012designing}, and shows that role-playing can support design processes \citep{simsarian2003take}. 
Building on these insights, participants first engaged in a role-play exercise in which one acted as a student and enacted a sample scenario by constructing an e-textile artifact using the ``SewSimple'' maker kit and BBC micro:bit to address a social problem. 

During workshop two, teachers collaboratively mapped which knowledge and skills in the scenario aligned with the National Curriculum for England (KS3) Computing and Design \& Technology (D\&T) goals. The aim was to show how a scenario can embody and integrate curriculum goals and to build the habit of explicitly referencing KS3 aims when designing their own scenarios later in the workshop. Five participants self-selected into two groups (in this instance, the participants self-selected working in one all-female, one all-male group) and worked on a pre-prepared whiteboard with the left-hand reference column listing the KS3 Computing and D\&T goals, and the right-hand column capturing the knowledge/skills evidenced by the scenario. Step by step, groups reviewed the scenario and identified a relevant KS3 Computing goal, then wrote the aligned knowledge and skills on sticky notes and placed them in the corresponding column. Next, building directly on the mapping outcomes, teachers refined their own SBL ideas for e-textiles and co-planned a lesson sequence that scaffolds the use of ``SewSimple'' and the BBC micro:bit. Using a provided flow-chart template, they documented the plan in a structured section, including activity description, required resources, and expected learning outcomes, therefore turning insights from enactment and mapping into tangible design artifacts ready for classroom use. Similar to the first workshop, the session concluded with group reflections with contrasting teacher-designed e-textile teaching plans to the KS3 computing education guidance \citep{govuk_k3_computing}. 

\begin{table}[!ht]
\centering
\caption{Participants' information. Some teachers (P1 and P4) participated in both workshops.}
\label{tab:workshop_participants}
\footnotesize
\setlength{\tabcolsep}{4pt}
\renewcommand{\arraystretch}{1.1}
\begin{tabularx}{\columnwidth}{@{} c c c l X @{}}
\toprule
\textbf{Workshop} & \textbf{ID} & \textbf{Gender} & \textbf{Subject} & \textbf{Level} \\
\midrule
W1 & P1 & M & Computing & Secondary \\
W1 & P2 & M & Computing & Secondary \\
W1 & P3 & F & Textiles & Secondary \\
W1 & P4 & F & Computing & Primary \\
W1 & P5 & M & Computing & Secondary \& Afterschool \\
W2 & P1 & M & Computing & Secondary \\
W2 & P4 & F & Computing & Primary \\
W2 & P6 & M & Computing & Secondary \\
W2 & P7 & F & Art \& Design & Primary \\
W2 & P8 & F & Textiles & Primary, Secondary \& Afterschool \\
\bottomrule
\end{tabularx}
\end{table}

\subsection{Data Collection and Analysis}
Both Workshops were facilitated by a graduate student with a background in learning experience design (LXD), supported by two authors (a faculty author and a PhD student) and another graduate student for data collection. Five hours of video data were collected along with tangible artifacts such as a teacher/participant-designed e-textile teaching plan.
We videoed all design activities and intra/inter-group conversations and used MS Teams to auto-transcribe the group discussion parts. For data analysis, we analyzed the transcriptions and design artifacts such as instructional materials for introducing e-textiles with the ``SewSimple''.

The team performed reflexive thematic analysis (RTA) \citep{BraunTerry2019Thematic, braun2021can, BruleThematicAnalysis2020, braun2023doing} due to its flexibility to combine the inductive coding process with practical pedagogical perspectives that drew on the coders’ lived experiences as former educators and learning technologists. The first author, a physical computing educator and one of the main creators of the ``SewSimple'', analyzed the workshop artifacts, while the second and third authors (who both are identified as female working in educational research and computer science) collaboratively coded two group discussions. In this process, we met regularly with the faculty authors on the team to discuss our initial codes and definitions, and refined our themes after revisiting the full dataset following the suggested six-phase approach \citep{braun2023doing}. 

\section{Results} 
\subsection{Opportunities}
Two themes emerged related to opportunities to integrate e-textiles into the English computing curriculum: foster students' motivation and interest, and a potential alignment with the English national curriculum. These are in line with the previous computing education research on leveraging physical computing's unique affordances to bridge the gap between code and the physical world into tangible, real-world experiences \citep{sentance2017b, Schmidt2016ComputerLiteracy}, particularly e-textiles that enable culturally relevant pathways to computing, improving students' self-perception of their STEM abilities \citep{kafai2014crafts, coulter2023textiles}. 

\subsubsection{Fostering Students' Motivation and Interest}
Teachers reported that multimodal experiences with physical computing technologies play a powerful role in sparking students’ motivation and sustaining their interest in computing practices. One teacher described how embedding a BBC micro:bit into an RC car transformed programming lessons into interactive experiences:\textit{``When the car moved, you could see the lights go on in their eyes, and they were genuinely trying to learn''}(P5). Physical computing systems afford a tangible way to deliver high-level computing concepts, through physical, real-world applications with Python coding exercises. It enhances young learners' engagement by making learning more concrete and rewarding. In contrast to knowledge mediated through screen texts, students saw the effects of their programming in the car's movement. E-textiles seemed to extend these benefits of physical computing, and offered a pathway to engage with students who are scared of traditional computing lessons: \textit{``I had a lot of issues with attracting young ladies to computer science, and I think that's (e-textile practices) a brilliant way to kind of entice them back''}(P2). Again, E-textiles can help reposition CS as interdisciplinary, connecting creativity, design, and engineering to challenge stereotypical perceptions of computing as purely technical and abstract, and attract broader participation. Taken together, these teacher's experiences illustrate how physical computing, especially e-textiles, can function as motivational catalysts to involve students with different experiences in learning. 

\subsubsection{Alignment and Diversifying Curriculum}
Teachers observed that e-textiles could easily be included in the national curriculum because their aims closely align with existing subject frameworks. One teacher (P3) reflected on the adaptability: \textit{``I feel like it definitely aligns with the curriculum. I feel like it has a lot of scope as well for quite interesting work...I mean, obviously I’m also in art and design.''} While working from a creative perspective, the teacher saw interdisciplinary potential to connect creativity and programming through making and design, bridging aesthetic exploration with technical skills. The computing dimensions of e-textiles reinforce its compatibility with national curriculum priorities for coding: \textit{``...with the micro:bit [as part of the e-textiles design], you’d link the coding part of it to the curriculum, so you can show that connection'' (P4).}  Although these opportunities could facilitate the integration of e-textiles into the curriculum, in practice, there are also some barriers that prevent the integration of e-textiles. 

\subsection{Barriers}
There are two main themes related to barriers to integrate e-textiles into English computing curriculum, including issues with teaching physical computing and curricular barriers.. 

\subsubsection{Issues with Teaching Physical Computing in Schools}
We identify two issues with teaching Physical Computing. First is the knowledge planning conflicts. Some of the teachers discussed the challenges of introducing BBC micro: bit and ``SewSimple'' in the current teaching practices at schools, particularly in the aspect of knowledge planning. Teachers noted that while these technologies have the potential to connect computing, electronics, and design, fragmented planning across departments (subjects) limited their impact and increased the cognitive demands placed on students to recall prior learning independently. For P2, they tried to work with the engineering department - but as the terms when he taught occurred did align, \textit{``by the time it came to spring, the students had already forgotten the `Python''' (P2).} There can be a mismatch in curriculum timing across the year, where students were expected to apply programming knowledge months after learning it, resulting in a significant loss of knowledge retainment. Teachers connected this issue to a broader aspiration for more coherent cross-subject planning:
\begin{quote}
Which [cross-subject planning] is what you would hope happens with the national curriculum—that we could link all of our subjects together in the same term and teach the same stuff. It doesn’t always happen that way (P4).
\end{quote}
 Such misalignment across subjects and terms created barriers for integrating e-textile activities effectively.

Second, there lacks sufficient time in the Formal Curriculum. Teachers reported the micro:bit was too time-consuming to use during the limited hours allocated for these lessons under the formal curriculum.  In particular, they were concerned about the time required to inventory, maintain, and distribute all of the requisite components. The ``SewSimple'' further exacerbated this due to the additional requirements of e-textile materials. Both P2 and P3 commented that the ``set-up'' required to lay out materials and equipment required for physical computing consumes excessive time and is ``annoying'' (P2): \textit{``10 minutes to get it out, 10 minutes to put it back in...the BBC [micro:bit] could have made it easier for the classroom...The cable is the barrier.}
P3 added: \textit{``If you’re teaching, like, five, two-hour lessons a day, like in PGCE art, you’re doing those setups all the time, and you feel like the actual teaching becomes a little bit harder.''} One teacher reflected on this significant time demands associated with delivering these activities as part of regular lessons rather than in an after-school club: \textit{``I took it out, like he (P1) said, because of the setup (P2)''.}  Therefore, due to the substantial time required for equipment setup, a large amount of lesson time was lost, leaving limited opportunities for actual teaching and learning activities. This time pressure made it difficult for teachers to meet curriculum objectives and maintain student engagement, leading to (P2) eventually removing these activities from the formal curriculum. The tension between the limited lesson time available and the practical requirements of teaching e-textiles highlights the need for adequate planning, scheduling, and resource management when introducing the BBC micro:bit, the ``SewSimple'' maker kit, or any other physical computing system. 

\subsubsection{Curricular Barriers in the UK Key Stage 3 Frameworks}
Participants particularly highlighted that disciplinary boundaries and self-contained subject-focused learning guided by the UK national curriculum hindered the integration of e-textiles into computing learning. For example, in the UK, Design \& Technology is considered as a closely relevant subject to e-textiles for computing education  \citep{coulter2023textiles}. While tools such as the ``SewSimple'' and BBC micro:bit are perceived aligning well with the aims of design technology (P3): \textit{``design technology is such a small part of the curriculum, and this would actually fit in really well, so yes, it does align with it''}, the subject itself occupies only one optional class under the national curriculum, resulting in a limited impact on leveraging the diversity goals of e-textiles. Furthermore, one teacher pointed out that the limited focus on physical computing and maker education in the national curriculum misses the opportunities to incorporate e-textiles more broadly into teaching practices and the school curriculum, which loses the opportunities to engage underrepresented groups, particularly girls in computing fields: 
\textit{``
And [I] think there should be more of a section on e-textiles...I think they've missed an opportunity there, and that will get more girls into actually ... using computers'' (P4).}
This was echoed by P1 and P2, who shared their struggles of including more girls in either school classes (P1) or technology clubs (P2) due to the fear of conventional computing education: \textit{`` The girls just think programming is hard, for whatever reason. But when they get it, when it clicks in their head, they fly; sometimes they fly faster than the boys.'' (P2)} Instead, P3 observed a gender imbalance in Art \& Design classes that \textit{``Interestingly, the class I taught for that was mostly girls and like less boys''}. She suggested that integrating e-textiles into the Art \& Design classroom practices could provide an opportunity to challenge the current gendered patterns of participation and become \textit{``an interesting way to get around''} to include more boys in creative skills development and support girls to develop computational interest extended from existing creative making learning. However, her comments revealed a new challenge: ensuring that e-textile activities are equally appealing to students of all genders and that engagement is sustained for all learners. 

These reflections underscore the tension between curriculum content and opportunities for inclusive computing education, illustrating that curriculum planning intentionally bridges design technology, art, and computing, and also promotes gender equity by providing a more diverse range of entry points into computing.

\subsection{Practical Challenges}
\textit{Learning Curve and Additional Support}.
Teachers reported that the ``SewSimple'' maker kit could extend BBC micro:bits' use in the current learning exercises; however, it faces several non-technology-related challenges for immediate classroom adoption. For example,  students lacked core crafting skills such as sewing required for all e-textiles. P4 highlighted that students could gradually build confidence in programming, another core knowledge area for e-textile making, through the MakeCode tutorials by BBC micro:bit education, while teachers \textit{``can spend more time on the textiles side, which is another challenge in itself, you know, sewing, using scissors, that kind of thing''.} She further noted in the second workshop when discussing e-textile class planning: \textit{``...you do have to teach the (crafting) skillsets. For example, some primary pupils couldn’t even thread a needle. So you can’t skip the foundational skills.''} This indicates a potential knowledge shift in focus for computing sessions to support successful and sustained learning with e-textiles. 

Similar to students needing to learn non-computing knowledge, another challenge involved teachers themselves acquiring new skills. As e-textiles enable cross-disciplinary teaching by combining Art \& Design with computing \citep{fields2017teaching}, teachers from each background also need a fundamental understanding of the other domain. For example, P3, an Art \& Design teacher, noted they \textit{``would struggle with the programming part''}. This highlights the need, particularly for less technical teachers, to build confidence and overcome the fear of learning to code. Similarly, P2, a secondary school computing teacher, expressed hesitation about lacking crafting knowledge and recognized that realizing the full potential of e-textiles in formal learning would require time for further exploration to ensure a smooth transition and sustainable use: \textit{``because textile is not my specialism. I feel like this kind of project will stay as a club for now, see how much time things take, what I need, that kind of thing, and then we could bring it into the classroom''}(P2).

Indeed, some of the computing teachers reluctantly conveyed they did not know how to sew. Overall, these findings show that beyond the technology platforms, successful adoption of e-textiles in computing will also require dedicated support for new knowledge development and time for experimentation for both students and teachers to help grow the cross-disciplinary skills and confidence needed to realize the full potential of e-textiles. 

\textit{Financial Costs of the Physical Materials}. \citet{rode2024spools} highlight that teaching computing with e-textiles is unique in that it leverages material qualities afforded by physical components such as the electronic components, conventional crafting supplies, and microcontrollers like the BBC micro:bit, and the integration of coding and crafting into interactive computing exercises that together prompt students to learn and apply computing knowledge to tangible learning outcomes. For teachers, this introduces a new challenge in this material exploration process: access to materials and the cost of damage. In the first workshop, participants shared their current teaching conditions at their schools and noted the need to purchase ``electronic basics'' to be able to afford e-textile sessions. As a newly qualified teacher, P3 added to the comment: \textit{``I just feel like, yeah, there are definitely lots of resources needed for it. I’m not sure how much that would cost, like thread and stuff, I don’t know how expensive that is.''} While having the demands, P1 pointed out ongoing budgeting issues at his school and across the nation: \textit{``a lot of design technology jobs are being cut...in my school and within the borough.''} These budget constraints also increased the concern around material supplies and costs; he highlighted that the wear and tear in physical computing projects makes it difficult to reuse limited resources: \textit{``imagine little kids put it, pulling out the battery pack, and there was a lot of damage there. So eventually we just had to kind of scrap it completely, which is a shame, because I would love to do something like that again''}.  This underscores the financial realities of material acquisition and maintenance are a significant barrier to the sustainable introduction of e-textiles and creative education in computing classrooms. 

\textit{Lack of Teaching Resources and Feasible Approaches for Teaching E-textiles}.
Three instances from Workshop 1's artifact (post-it note) analysis illustrated video demonstrations as a teaching method and highlighted the need for further support to effectively introduce e-textile knowledge with the maker kit. This discussion continued in Workshop 2, where four teacher participants mentioned the potential suitability and usefulness of video for teaching current students: \textit{``I think in today's age, I think a lot of the kids get things quickly from animation, video, animations. I mean, that's it. They run with it. They're like, okay, I get it. And off they go''}(P7). P8, however, expressed some concerns, noting \textit{``for some students, it needs to be slightly tailored, because not all of them actually understand what you mean right away.''} This shows that the weakness of video demonstrations lies in their lack of individual, personalized focus, which contrasts with the strengths of e-textile computing education as discussed in other literature \citep{fields2023supporting}. Furthermore, the effectiveness of such an approach was questioned in the first workshop. P2 shared his observation of the official website’s instructions and students' learning experiences. However, one also raised questions about how to plan step-by-step content that would be suitable and easier for students to understand. He further said, \textit{``I think it really tricks the girls, because on the BBC website, did you push the fact you can sew it to things like that? But it feels like an afterthought.''} This shows that while existing online e-textile content may attract learners of a wider range of genders, transforming this interest into actual engagement through video tutorials alone is not enough. Overall, these reflections demonstrate that video-based learning may spark initial curiosity. However, its ``one-size-fits-all'' format may limit deeper engagement, highlighting the need for instructional materials and methods sensitive to diverse learner needs.


\section{Discussion}
\subsection{Curriculum Gaps and Opportunities}
We found that there are structural challenges of integrating e-textiles into the KS3 computing curriculum. The current national curriculum provides very limited coverage of physical computing, and as \citet{rode2024exploring} note, the design of the BBC micro:bit does not fully account for gender equity or integration with e-textiles. The lack of curricular grounding leaves tools such as ``SewSimple'' without a formal entry point. Teachers repeatedly reported that even when students had previously been introduced to Python, the absence of continuity between subjects meant that knowledge was often forgotten or difficult to transfer. P2 explained, they taught Python in the autumn term, but by the time they collaborated with engineering in the spring, the students had already forgotten it. These down stream effect exemplifies the consequences of weak curriculum integration and reinforces \citet{sentance2017b}'s argument that micro:bit teaching requires adequate scaffolding and professional development.

The study also demonstrates that implementing e-textiles requires teachers to master programming, electronics, textile craft skills, and interdisciplinary pedagogy simultaneously. This is not only a matter of acquiring additional skills but represents a reconfiguration of professional identity \citep{videnovik2018}. Traditionally, computing teachers have been positioned as “coding experts,” while D\&T or art teachers have been seen as “craft and design experts.” Electronics skills are often outside both roles professional senses of self.  By contrast, our participants revealed the challenges of occupying cross-disciplinary roles. P3 admitted that they would struggle with the coding part, while P2 stated that textiles is not their domain. This extends \citet{fields2017teaching}’s work on how e-textiles foster personalization and collaborative learning, offering a new perspective on how teachers adapt during processes of technological integration.

Yet, the UK context still lacks robust systems of teacher support. Despite efforts, the government-funded computing resource such as the National Center for Computing Education \citep{national_computing}, which provides discipline-specific support, only provides minimal content for the BBC micro:bit and none for e-textiles \cite{wohl2025learning}. Compared to the US, where teacher communities and structured professional development in e-textiles are more established \citep{fields2023supporting}, UK teachers reported isolation, lack of time for self-study, and minimal access to shared resources. As \citet{boulton2018implementing} argue, while some communities of practice and regional hubs have emerged, their reach remains uneven and sustainability uncertain. Without systemic support at the national level, e-textiles are unlikely to be sustainably embedded in the formal curriculum. At the same time, the findings confirm the unique educational value of physical computing. Teachers consistently highlighted that combining micro:bit with ``SewSimple'' promoted tangibility, motivation, collaboration, and creativity, offering clear advantages over purely software-based programming. This aligns with findings that 86\% of students reported computing as more enjoyable, with girls 70\% more likely to consider computing as a subject choice \citep{austin2020bbc}. Building on the computational making framework \citep{rode2015computational}, our study shows ``SewSimple'' demonstrates the feasibility of cross-disciplinary learning within existing infrastructures. However, its sustainability remains dependent on curriculum structures and policy support.

\subsection{Gender, Curriculum Silos, and Inclusivity}
Our study further uncovers the relationship between curriculum silos and gender bias. The 2014 reform of the UK curriculum \citep{govuk_k3_computing} emphasized computer science while neglecting integration with arts and design \citep{wohl2025learning, coulter2023textiles}.  This shift required teachers to move from teaching basic ICT skills to programming, algorithms, and computational thinking \citep{wohl2025learning}. The result, as demonstrated by both the literature and our data, is that computing lessons are broadly dominated by boys, while Art \& Design classes continue to primarily attract more girls. This structural division reflects a gap in curriculum design that reinforces gender bias in computing as well as other subjects. As \citet{sentance2017computing} report, many teachers highlight disengagement—particularly among girls. This was echoed in our study by P4's point that there’s not enough access to physical computing or e-textiles, which is a missed opportunity to get more girls truly using computing.

E-textiles provide a means of bridging disciplinary divides. Our teachers observed that girls often reported low self-efficacy in programming, which is consistent with previous work on gender and programming \citep{kallia2018boys, aivaloglou2019early}. P1 noted that girls just find programming hard, but when they get it, and it clicks in their head, they fly, and sometimes even faster than the boys. Activities with ``SewSimple'' enabled students to achieve tangible mastery experiences, aligning with international evidence that, under targeted encouragement, girls not only match but may even outperform boys in programming tasks \citep{videnovik2018}. Such potential is emphasized by \citet{tzagkaraki2023}, who found that micro:bit-based activities significantly enhanced students’ self-efficacy and career motivation. Several teachers remarked that boys, through participation in sewing and crafting tasks, engaged with creative practices. P3 described this as an interesting way to ``get around'', which enables girls to build interest in computing through creative making, while encouraging boys to value design and craft. This dual opportunity is underrepresented in the literature and provides an important theoretical contribution.

Prior work highlights broader inclusivity potential of e-textiles such as fostering welcoming classroom environments \citep{nakajima2019teachers} and making computing meaningful for all learners \citep{shaw2020leveraging}. Yet our findings show that current curriculum, which is narrowly focused on programming skills in its delivery, limits e-textiles' ability to realize these diversity goals. As P4 observed, D\&T is such a small part of the curriculum. Although the SewSimple fits really well, its marginal status undermines its impact. These results underscore the need to revisit the philosophical foundations of the national curriculum. Rather than positioning computer science as the exclusive center, a more balanced, interdisciplinary, and gender-inclusive approach is required to ensure equitable participation for all students.

\subsection{Funding, Policy, and Partnerships}
Lastly, our study discussed how limited school budgets, which are part of an ongoing financial issue across UK primary and secondary schools \citep{neu_budget}, have become a key constraint on the smooth integration of e-textiles and physical computing in secondary computing classrooms. Previous research by the Raspberry Pi Foundation, a UK-based independent educational charity, found that although physical computing technologies, particularly the BBC micro:bit, have been given governmental support for computing education, many English primary school teachers still found it struggling to secure sufficient equipment for classroom teaching and faced inequities in access due to factors such as school size, region, and category (independent v.s. state-funded) \citep{raspberrypi_report}. In the e-textile context, our findings echoed this discussion, highlighting that resource shortages raised teachers' concerns to obtain enough a wide range of materials required for the material-driven practices unique to e-textile learning \citep{buechley2007towards, rode2024spools}. Further, our findings show that this financial issue also hinders having qualified e-textile teachers in the classroom due to staff recruitment or limited support for teachers’ skill development across cognate subject areas such as Design \& Technology or Art \& Design \citep{coulter2023textiles}—creative disciplines core to developing fundamental e-textile skills such as creativity and constructing \citep{rode2015computational}. 

While prior work has not directly emphasized these financial barriers to teachers' experiences with e-textiles in CS education, it has provided insights into best practices to support teacher learning and managing material concerns. For example, \citet{fields2017teaching} demonstrated the success of collaborations between research institutions and schools through the government-supported program ``Exploring Computer Science'', in which computing teachers worked with e-textile scholars and practitioners to develop core skills and teaching materials suitable for e-textile use in the computing classroom. In the UK, several universities—including the institution where this study was conducted—offer widening-participation programs (e.g., the CS4fn project by Queen Mary University of London) as an outreach initiative to foster collaboration with schools and individuals \citep{royalsociety_report}. While these initiatives may not directly resolve funding issues, they can create an opportunity and sustain teacher and student engagement with physical computing equipment and e-textiles, paving the way to future e-textile curricular implementation. 

Furthermore, \citet[p. 13]{shaw2020leveraging} emphasized the benefit of leveraging local resources to enable transforming existing knowledge to effective e-textile teaching in schools and facilitate teachers' ``grew (growing) existing relationships and developed new ones with each other'' through participatory, community-based learning environments \citep{fields2023supporting}. Some British after-school programs focused on computing and technology, such as Code Club and CoderDojo, operate on a volunteer-based model and provide extracurricular knowledge and skills to support students’ informal learning and help them maintain technological skills relevant to the digital economy \citep{royalsociety_report}. Although teachers may not be directly involved, many teachers like some of our participants reported prior experience working in these clubs and building new relationships that help knowledge exchange between school and informal learning environments. Such teacher communities are critical for enabling cross-subject and cross-disciplinary collaboration \citep{fields2017teaching} but also afford social connections and emotional support for continuous teacher's development \citep{fields2023supporting}, both of which are important to integrate e-textiles into computing teaching practice at schools. Overall, these communities may provide a foundation for building sustainable networks that support both professional development and resource sharing for e-textile computing, ultimately benefiting both teachers and students.

\subsection{Limitations}
One study limitation is that due to the physical locations' constraints, our participants are mainly recruited from London-based schools, which may restrict our understanding of a wider range of teachers' and students' experiences with educational technologies, particularly those from areas and regions where technology resources are more scarce. Furthermore, our study was grounded in an analysis of the English National Curriculum, which limits our discussion of educational objectives outlined by other nations in the UK. Although our findings are context-specific, prior research indicates substantial similarities across K–12 CS education internationally \citep{falkner2019international}. We therefore believe that our insights have broader relevance for curricula where subjects are siloed. At the same time, we call for more region-specific research to better leverage the educational potential of e-textiles for both teachers and students. Finally, we acknowledge that our research and much prior work were conducted in a binary gender context, and additional research in a gender-diverse context is needed to be more inclusive.

\section{Conclusion and Future Work} 

Physical computing technologies such as the BBC micro:bit and e-textiles have shown great potential in supporting equity goals in secondary school computing education. Through two participatory design workshops with eight teachers, we identified opportunities that e-textiles could foster students' motivation and interests in learning through real-world and physical application of computing, and that e-textiles align with some of the aims of the national curriculum. However, teachers also reported persistent barriers, including fragmented cross-subject planning and limited curriculum time, the lack of physical computing in the national curriculum, practical challenges such as complex classroom setup and new skill demands (both for students and teachers), financial constraints around materials, and a lack of tailored teaching resources. These structural and practical obstacles highlight the need for clearer curriculum design, affordable and reusable materials access, and professional development to build teachers’ confidence in delivering e-textile activities effectively. More importantly, the national curriculum needs to be revised to enable teaching across all three areas, including programming, electronics, and e-textiles. At the same time, the findings highlight structural tensions that shape classroom realities. While teachers perceived strong curricular alignment between e-textiles and the national computing framework, fragmentation across departments and rigid curriculum structures limited coherent cross-subject planning. These perceptions reveal a tension and broader systemic pressures facing UK schools: e-textiles can make computing more tangible, collaborative, and inclusive, yet systemic and institutional factors—curriculum silos, resource scarcity, and limited training—currently inhibit their impact. This research was one of the few that uses a purpose-designed e-textile device to explore the use of this approach with teachers; rather than highlighting the hardware or programming environment barriers, this study was able to focus on the practical challenges of including the e-textile approach in UK classrooms.  This research contributes to computing education in three key ways. by examining teachers’ direct, cross-disciplinary experiences with scenario-based learning using a newly designed e-textile platform, identifying opportunities, barriers, and challenges, and finally offering insights into best practices for supporting e-textile adoption and pedagogical practices within existing institutional structures to maximize students' benefits for secondary school computing education.

Beyond these curricula concerns, while we have a lot of e-textiles research focusing on attracting girls to computing \citep{buechley2008lilypad, ioannou2025learning, feng2025sewsimple}; less is known about boys' attitudes towards art and crafting skills development. Certainly, our data shows these activities are perceived as highly gendered. \citet[p.15]{weibert2014sewing} were careful to present data to push back that boys can find e-textiles fulfilling, writing that the use of an e-textile \textit{``allows both girls and boys to demonstrate technical mastery as well as to explore and construct a spectrum of gendered sociotechnical identities that might otherwise be obscured by conventional masculinist attitudes towards technology''}. Elsewhere, \citet{rode2025reframing} have discussed that there are pressures on technology learners to perform their gender in line with social norms about appropriate technology use by their gender (this is based on the social theory of ``technology as masculine culture'' \citep{wajcman1994technology}).  As e-textiles are highly gendered, this will remain a significant challenge for teachers; consequently, future work must unpack not only how to support the development of self-efficacy but also how teachers actively teach to decouple love of technology, including e-textiles, from pervasive gender norms. 

To support teachers, we highlighted ways to form and strengthen teacher networks and to foster a community that supports teachers and students' engagement with e-textiles and other physical computing technologies. Based on the UK context, we specifically identified opportunities such as collaborations between research institutes and schools or interested individuals, as well as connections with volunteer tutors from government-funded after-school clubs. These efforts can facilitate skill exchange, resource sharing, and continuous learning that help both students and teachers prepare for future policy changes. 

\begin{acks}
This work was partially funded by the UCL Higher Education Innovation Fund (HEIF) Knowledge Exchange and Innovation program (KEI2024-01-65). We particularly thank Claire Qiu for supporting data collection, and Andrian Mee and Carol Wild for recruitment. 
\end{acks}

\bibliographystyle{ACM-Reference-Format}
\bibliography{reference}

\end{document}